\documentclass[
reprint,
longbibliography,
aps,pra,
twoside,
bibnotes,
floatfix
]{revtex4-2}
\usepackage{tabularx}
\usepackage{physics}
\usepackage{dsfont}
\usepackage{color,newfloat}
\usepackage{graphicx}
\usepackage{dcolumn}
\usepackage{bm}
\usepackage{amsmath,amssymb,amsthm,amsfonts}
\usepackage{mathtools}
\usepackage{multirow}
\usepackage{hyperref}
\usepackage[all]{hypcap}
\usepackage{chngcntr}
\usepackage{enumerate}
\usepackage{lipsum}
\usepackage{mathrsfs}
\usepackage[none]{hyphenat}

\newcommand{\comment}[1]{}

\begin{document}
\title{Experimental entanglement generation using multiport beam splitters}

\author{Shreya Kumar$^{1,2}$, Daniel Bhatti$^{1,2}$, Alex E. Jones$^3$, and Stefanie Barz$^{1,2}$}
\affiliation{
$^1$Institute for Functional Matter and Quantum Technologies, Universit{\"a}t Stuttgart, 70569 Stuttgart, Germany \\
$^2$Center for Integrated Quantum Science and Technology (IQST), Universit{\"a}t Stuttgart, 70569 Stuttgart, Germany\\
$^3$Quantum Engineering Technology Labs, H. H. Wills Physics Laboratory and Department of Electrical and Electronic Engineering, University of Bristol, Bristol BS8 1FD, UK}

\begin{abstract}

Multi-photon entanglement plays a central role in optical quantum technologies. 
One way to entangle two photons is to prepare them in orthogonal internal states, for example, in two polarisations, and then send them through a balanced beam splitter.
Post-selecting on the cases where there is one photon in each output port results in a maximally entangled state.
This idea can be extended to schemes for the post-selected generation of larger entangled states. Typically, switching between different types of entangled states require different arrangements of beam splitters and so a new experimental setup.
Here, we demonstrate a simple and versatile scheme to generate different types of genuine tripartite entangled states with only one experimental setup. We send three photons through a three-port splitter and vary their internal states before post-selecting on certain output distributions. This results in the generation of tripartite W, G and GHZ states. We obtain fidelities of up to (87.3$ \pm 1.1)\%$ with regard to the respective ideal states, confirming a successful generation of genuine tripartite entanglement.
\end{abstract}

\maketitle
\section{Introduction}

Multipartite entangled states enable networked quantum communication protocols, such as networked quantum key exchange, quantum secret sharing, 
and measurement-based quantum computation~\cite{murta2020,portmann2022}. 
The type of entangled resource required depends on the particular task. For instance, the maximal quantum correlations present in Greenberger-Horne-Zeilinger (GHZ) states make them ideal for quantum communication protocols, whereas W states exhibit multipartite entanglement even after photon loss~\cite{murta2020}. Also, G states, which have similar loss robustness as W states, have been introduced in the context of quantum secret sharing~\cite{Sen2003}.

On the experimental side, many approaches to generate different types of entangled states using photons have been realized~\cite{eibl2004,kiesel2007,zhong2018,thomas2022,lee2022,meyer-scott2022}. Using post-selection maximally entangled two-photon Bell states are relatively simple to produce~\cite{Shih1988,kwiat1995}.
However, as the number of photons increases, the generation of post-selected genuine multipartite entangled states becomes more challenging due to the complicated experimental setups, higher sensitivity to imperfections, and lower success probabilities~\cite{zhong2018,meyer-scott2022}.
For example, four-photon W states and three-photon G states have been generated using post-selection~\cite{prevedel2009, kiesel2007}, and post-selected GHZ states have been generated experimentally with up to 12 photons, and with 18-qubits encoded in 6 photons~\cite{zhong2018,wang2018}.
Besides post-selection, a different approach would be to herald the generation of multipartite entangled photon states measuring additional ancilla photons~\cite{walther2007}. So far, the experimental generation of heralded entangled photon pairs has been demonstrated~\cite{barz2010,wagenknecht2010}.

In these experiments, the setups were built and optimised for the generation of a specific quantum state~\cite{eibl2004,zhong2018,wang2018,lee2022,meyer-scott2022}. This means that switching between generation of, for example, GHZ and W states, would require modifications to the experimental setup~\cite{lee2022}. Here, we show that the same experimental setup can be used to generate GHZ, G and W states, using independent photons, a multiport splitter, and post-selection on certain photon number output distributions.

Our experiment uses photons at telecommunication wavelengths and so generate states suitable for use in networked quantum protocols using existing fibre networks or in the quantum internet~\cite{Wehner2018}. The setup presented in this work can be used as a central quantum server, providing resource states to different parties in a network.
In this work, we focus on states of three qubits, but our scheme can be extended to arbitrary size W and GHZ states as shown in~\cite{Lim2005,Bhatti2022}. The extension of the generation of G states is currently an open research question.

\section{Theory}
\label{sec:Theory}

In this work, we are interested in tripartite entangled states; for this case there exist two different classes of genuine entanglement: the class of W states and the class of GHZ states~\cite{dur2000}.
While states from the same class can be transformed into each other by means of local operations and classical communication, states from the two different classes cannot~\cite{dur2000}.
The two canonical tripartite entangled states are~\cite{dur2000,greenberger1990}:
\begin{align}
        \ket{\text{W}} &= \frac{1}{\sqrt{3}}\left(\ket{001}+\ket{010}+\ket{100}\right)  , \label{eq:W-state} \\
				\ket{\text{GHZ}}&=\frac{1}{\sqrt{2}}\left(\ket{000}+\ket{111}\right),
\end{align}
where $\ket{\psi_{1}\psi_{2}\psi_{3}} = \ket{\psi_{1}} \otimes \ket{\psi_{2}} \otimes \ket{\psi_{3}}$ with $\ket{\psi_{i}}=\ket{0},\ket{1}$ ($i=1,2,3$). 
Another highly entangled quantum state, which belongs to the GHZ-state class~\cite{dur2000,coffman2000}, is the tripartite G state. It is given by~\cite{Sen2003}
\begin{equation}
	\ket{\text{G}}=\frac{1}{\sqrt{2}}\left(\ket{\text{W}}+\ket{\overline{\text{W}}}\right),
\end{equation}
where
\begin{equation}
\ket{\overline{\textrm{W}}}=\frac{1}{\sqrt{3}}\left(\ket{110}+\ket{101}+\ket{011}\right).
\end{equation}

\begin{figure}[t]
    \centering
    \includegraphics[width=0.7\linewidth]{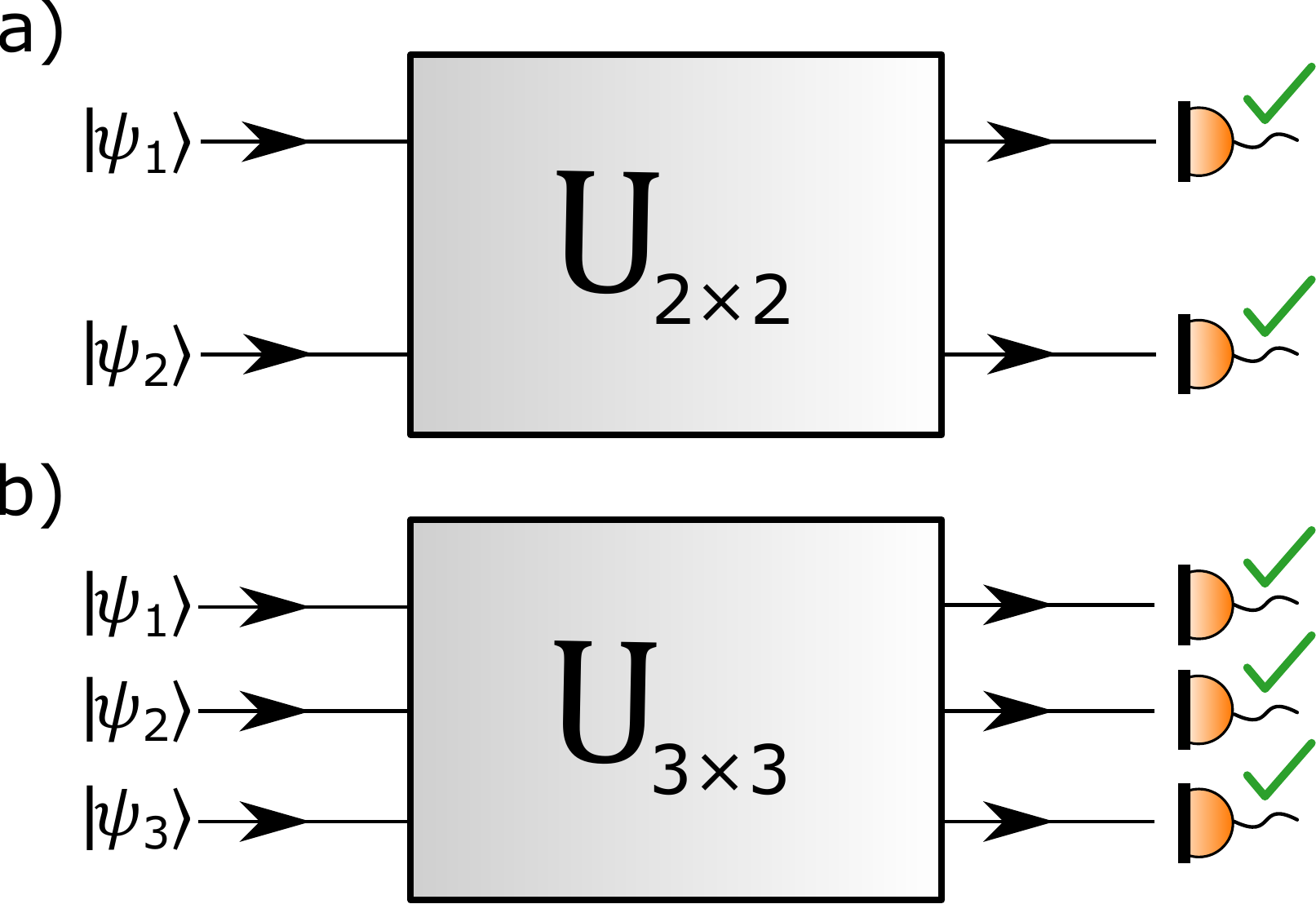}
    \caption{Schematic of a balanced (a) $2\times2$ and (b) $3\times3$ beam splitter, which can be used to generate bi- and tripartite entangled states, respectively. Detection of one photon per output mode indicates a successful generation of the entangled state. }
    \label{fig:23portBS}
\end{figure}
In order to generate multipartite entangled quantum states using post-selection, we focus on the interference of $N$ independent photonic qubits passing through a balanced beam splitter. A balanced beam splitter evenly directs $N$ spatial input modes to $N$ spatial output modes. The (canonical) discrete Fourier transform interferometers are described by the unitaries~\cite{Lim2005}:
\begin{equation}
    U_{kj}=\frac{1}{\sqrt{N}} \omega_N^{(k-1)(j-1)},
    \label{eq:matrixelements}
\end{equation}
where $\omega_{N}=\exp(2\pi i/N)$, and $k \textrm{ and } j$ are the indices of the input and output port, respectively. Let us now assume that the $N$ photons are prepared in a separable pure state of the form
\begin{align}
		\ket{\Psi}=\ket{ \Psi_{1},\ldots,\Psi_{j},\ldots,\Psi_{N}},
\end{align}
where $\ket{\Psi_{j}}$ denotes the internal state---in our case, polarisation---of a photon in the $j^{\textrm{th}}$ input port of the symmetric beam splitter. Here, the input and output ports, that is, the spatial modes of the photons, are the external degree of freedom.

The initial state evolves through the interferometer, which affects their spatial modes but leaves the polarisation of the photons unchanged. This leads to a big entangled state at the output containing many different terms, where each term has a certain number of photons in each output port, regardless of the internal state. Post-selecting for $N$-fold coincidences from the complete output state at the $N$ outputs of the symmetric beam splitter finally generates specific entangled $N$-partite states. The form and the type of entanglement depend on the internal degrees of freedom of the input photons~\cite{Lim2005,Bhatti2022}.

\begin{figure}[b!]
	\centering
	\hspace*{-0.25cm}
	\includegraphics[width=1.00\linewidth]{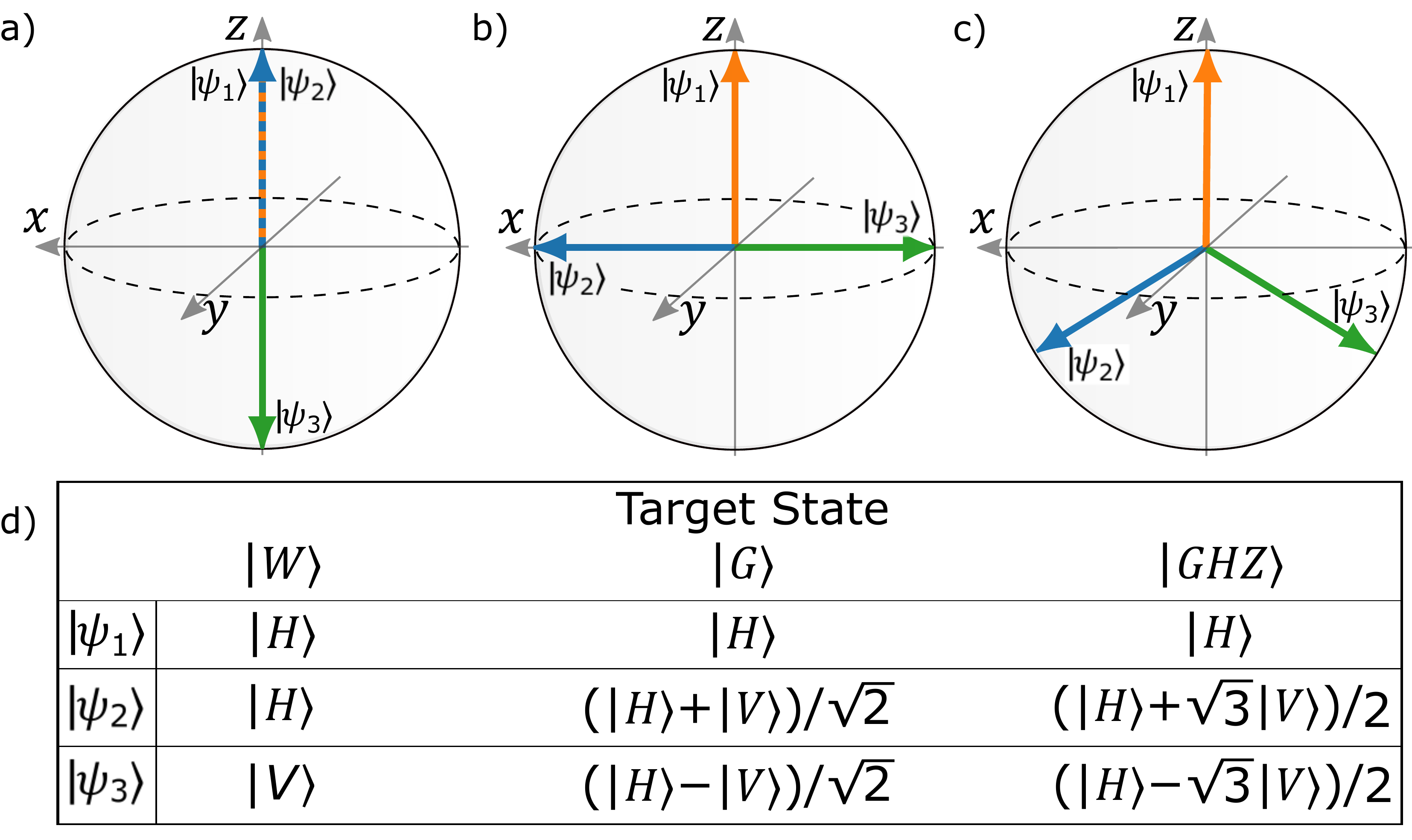}
	\caption{(a--c) Bloch-sphere representation of the input  states for W, G$^\prime$ and GHZ$^\prime$ state generation, respectively. (d) Polarisations of the three independent input photons that result in the generation of a polarisation-encoded states $\ket{\text{W}}$, $\ket{\text{G}^\prime}$ and $\ket{\text{GHZ}^\prime}$ via post-selection.}
	\label{fig:input states}
\end{figure}

For $N=2$, we have a balanced beam splitter with two input and output modes as shown in Fig.~\ref{fig:23portBS}(a). For instance, consider two orthogonally polarised photons, ${\ket{\Psi_{1}}=\ket{H}\equiv\ket{0}}$ (horizontal polarisation) and ${\ket{\Psi_{2}}=\ket{V}\equiv\ket{1}}$ (vertical polarisation). Post-selecting for two-fold coincidences at the output ports results in a maximally entangled two-photon state~\cite{Shih1988}
\begin{align}
 \ket{\Psi_{\text{out}}} = \frac{1}{\sqrt{2}} \left( \ket{HV} - \ket{VH} \right),
\end{align}
with a success probability of $50\%$.
This way of generating bipartite entanglement is well-known and has experimentally been verified and used in various applications~\cite{kim2001,wei2020}.
\begin{figure*}[!th]
    \centering
    \includegraphics[width=1\linewidth]{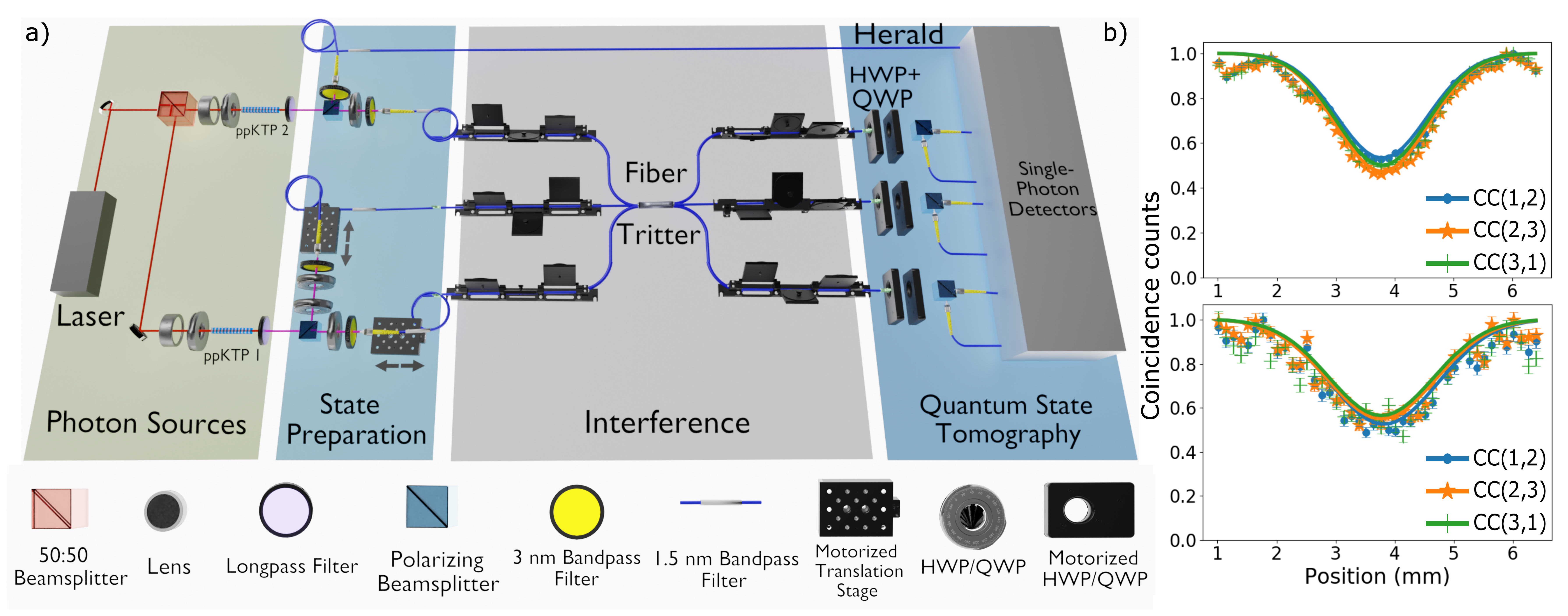}
    \caption{(a) Overview of the experimental setup: The photon sources are based on type-II spontaneous parametric down conversion in periodically poled titanyl phosphate crystals (ppKTP,  $1~\textrm{mm} \times 1~\textrm{mm} \times 30~\textrm{mm}$). Each source is pumped by a picosecond pulsed laser at $775~\textrm{nm}$ with an average power of $50~\textrm{mW}$, generating two photons at $1550~\textrm{nm}$.
		The generated photon pairs are orthogonally polarised and are separated using polarising beam splitters (PBSs).	The polarisation states of the photons are set using half-wave plates (HWPs) and quarter-wave plates (QWPs) and their temporal delays are controlled through linear translation stages, before being coupled into the tritter. The fourth photon is used as a herald.
     Quantum  state tomography is performed using wave plates and polarising beam splitters. The photons are detected with superconducting nanowire single-photon detectors (efficiencies $> 90\%$). 
    (b) HOM interference between signal and idler photons from source 2 (top) with an average visibility of $(49.94\pm0.02)\%$, and heralded HOM interference between idler photon of source 1 and signal photon of source 1 (bottom) with an average visibility of $(47.80\pm0.03)\%$. The maximum visibility for two indistinguishable photons at the inputs of a tritter is $50\%$.  CC(x,y) refers to coincidence counts between outputs x and y of the tritter, with x,y $\in\{1,2,3\}$. The error bars shown are $\sqrt{N}$, arising from the Poissonian nature of the photon counting process. The data was fitted with a Gaussian distribution (solid line).}
    \label{fig:setupoverview}
\end{figure*}

Scaling up to $N=3$, we have a balanced triport beam splitter, a so-called tritter, as shown in Fig.~\ref{fig:23portBS}(b). 
The increased system size allows post-selected generation of the different tripartite entanglement classes.
By employing the transformation matrix of a tritter~\cite{Zukowski1997},
\begin{equation}
U_3
=\frac{1}{\sqrt{3}}
\begin{pmatrix}
1 &1 & 1 \\
1 & e^{\frac{2\pi i}{3}}& e^{\frac{4\pi i}{3}} \\
1 & e^{\frac{4\pi i}{3}} & e^{\frac{8\pi i}{3}}
\end{pmatrix},
\label{trittermatrix}
\end{equation}
and the input  states listed and depicted in Fig.~\ref{fig:input states}, we 
post-select on three-fold coincidences to generate the following tripartite entangled states:
\begin{align}
     \ket{\text{W}}&=\frac{1}{\sqrt{3}}(\ket{HHV}+\ket{HVH}+\ket{VHH}), \label{eq:W state}\\
\begin{split}
\ket{\text{G$^\prime$}}&=\frac{1}{2\sqrt{3}}(3\ket{HHH}-\ket{HVV}-\ket{VHV}-\ket{VVH}), \label{eq:G state}
\end{split}\\
\begin{split}
    \ket{\text{GHZ$^\prime$}}&= \frac{1}{2}(\ket{HHH}-\ket{HVV}-\ket{VHV}-\ket{VVH}). \label{eq:GHZ state}
\end{split}
\end{align}
The states $\ket{\text{G$^\prime$}}$ and $\ket{\text{GHZ$^\prime$}}$ correspond to G and GHZ states in the X and Y basis, respectively.
These can be transformed into the states $\ket{\text{G}}$ and $\ket{\text{GHZ}}$ by applying single-qubit unitaries to each qubit~\footnote{The two unitaries that need to be applied to each qubit in the states $\ket{\text{G$^\prime$}}$ and $\ket{\text{GHZ$^\prime$}}$ to transform them into the states $\ket{\text{G}}$ and $\ket{\text{GHZ}}$ are given by:
\begin{align*}
    U_{\text{G$^\prime$}} = H = \frac{1}{\sqrt{2}}\begin{pmatrix}
1 & 1 \\
1 & -1
\end{pmatrix},
\end{align*}
and
\begin{align*}
U_{\text{GHZ$^\prime$}} = H \begin{pmatrix}
1 & 0 \\
0 & i
\end{pmatrix} 
= \frac{1}{\sqrt{2}}\begin{pmatrix}
1 & i \\
1 & -i
\end{pmatrix},
\end{align*}
where $H$ denotes the Hadamard matrix}.
The probabilities for successfully post-selecting the respective quantum  states are $p_{\text{W}}=p_{\text{G$^\prime$}}=1/9$, and $p_{\text{GHZ$^\prime$}}=1/12$.

\section{Experiment and results}
Here, we demonstrate the generation of the three different multipartite entangled states given by Eqs.~(\ref{eq:W state}--(\ref{eq:GHZ state}) in one experimental setup. Three photons pass through a fibre-based tritter and post-selecting one photon in each output mode of the tritter results in a tripartite entangled state  (see details in Fig.~\ref{fig:setupoverview}). 
\begin{figure*}[ht!]
    \centering
    \includegraphics[width=1\linewidth]{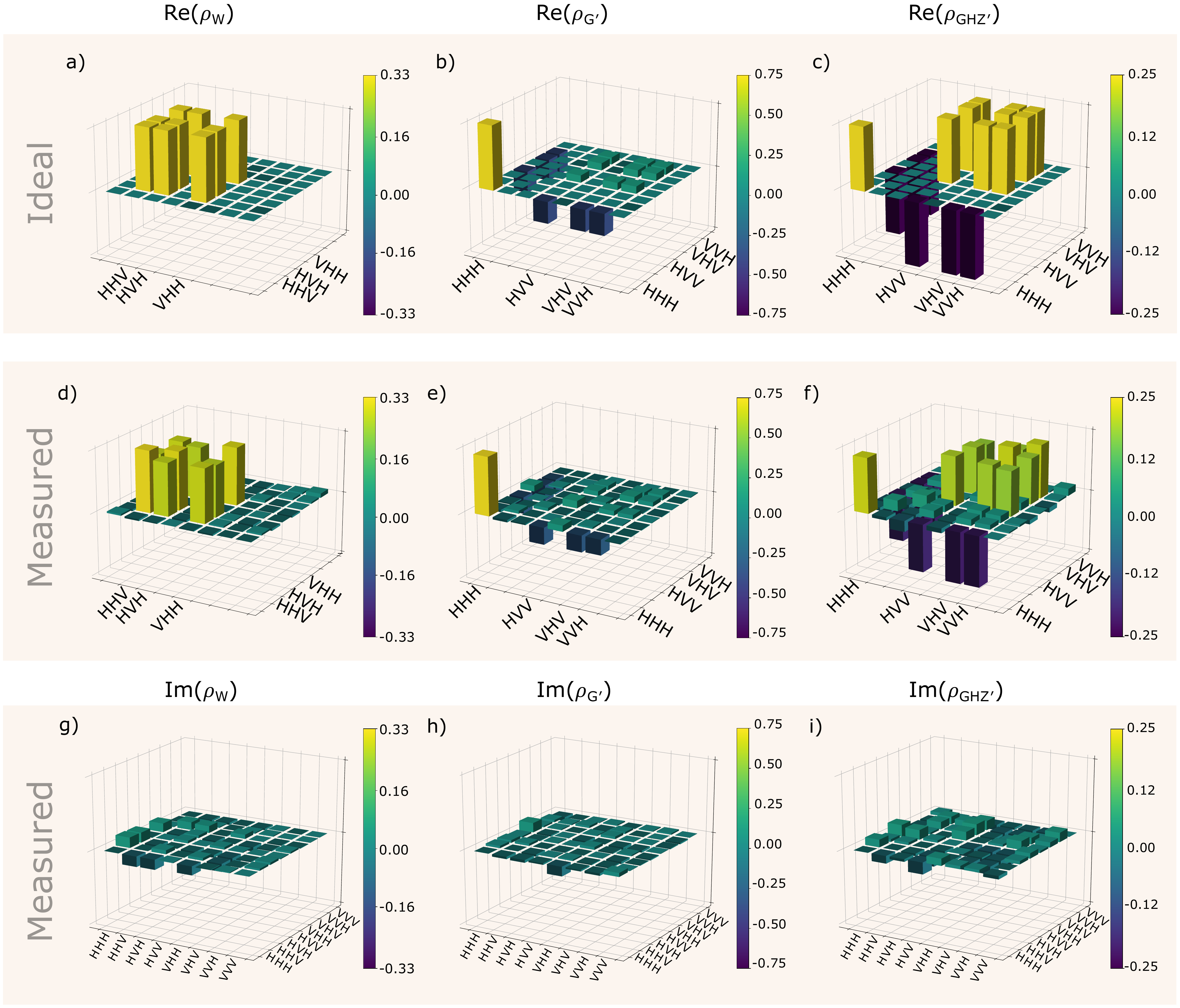}
    \caption{(a-c) Ideal density matrices of the states $\ket{\text{W}}$, $\ket{\text{G}^\prime}$, and $\ket{\text{GHZ}^\prime}$, respectively, where the dominant components are labelled on the axes. (d-f) Real part of the reconstructed density matrices 
    	with fidelities ${F_{\textrm{W}}=0.873 \pm 0.011}$, $F_{\textrm{G}^\prime}= 0.834 \pm 0.011$, and $F_{\textrm{GHZ}^\prime}= 0.788 \pm 0.016$, respectively. (g-i) Imaginary part of the reconstructed density matrices.
		}
   \label{fig:Tomography}
\end{figure*}

First, we characterise the tritter. We send in classical laser light at $1550~\textrm{nm}$ through each of the input ports, individually. The intensity of the light is measured at each of the outputs, which gives us the splitting ratios of the device. These splitting ratios also give us the absolute values of the elements comprising the unitary that describes the tritter operation. The splitting ratios along with the measured insertion losses are presented in Table \ref{tab:power}.
\begin{table}[b]%[h!]
    \centering
    \begin{tabularx}{8.28cm}{|p{1.5cm}|p{1.5cm} |p{1.5cm} |p{1.5cm}|p{1.5cm}|}
    \hline
         & Output 1 (\%)& Output 2 (\%)& Output 3 (\%)& Insertion loss (dB)\\
         \hline
        Input 1 & 32.01 & 30.24 & 29.86 & 0.356\\
        \hline
        Input 2 & 33.05 & 29.18 & 29.75 & 0.363 \\
        \hline
        Input 3 & 32.97 & 27.92 & 29.94 & 0.409\\
        \hline
    \end{tabularx}
    \caption{Measured splitting ratio and insertion loss of the tritter}
    \label{tab:power}
\end{table}
From this, we determine the normalised magnitudes of the elements of the transfer matrix of the tritter as:
\begin{equation}
    U_{\text{tritter}}^{\text{Exp}}= \frac{1}{\sqrt{3}}\begin{pmatrix}
    0.987 & 1.02 & 0.997\\
    1.00 & 0.999 & 0.997\\
    1.01 & 0.984 & 1.01\\
    \end{pmatrix},
\end{equation}
using the Sinkhorn-Knopp algorithm~\cite{sinkhorn1967}. The polarisation dependency of the device was checked by sending in horizontally and diagonally polarised light. The splitting ratio of the device was found to be similar for both polarisations, with a percentage difference of $2.3\%$ between them.\\
To test the distinguishability of the photons, we perform Hong-Ou-Mandel (HOM) interference experiments. The results are shown in Fig.~\ref{fig:setupoverview}b). We obtain an average visibility of $(49.94\pm0.02)\%$ for photons emitted from one source, and an average visibility of $(47.80\pm0.03)\%$, for heralded photons from different sources. Here, the visibility is calculated by ${V=(N_{\textrm{max}}-N_{\textrm{min}})/N_{\textrm{max}}}$, where $N_{\textrm{max(min)}}$ is the number of maximum (minimum) counts measured. Note that for a tritter with an ideal transfer matrix as given by Eqn.~\ref{trittermatrix}, the probability of observing two-fold coincidences between any two output ports is ${P_{11}=(2-|\braket{\psi_{j}}{\psi_{k}}|^2)/9}$. This results in a maximum visibility for two indistinguishable photons at any two input ports to be $50\%$. 

Now we set the polarisations of the three input photons according to Fig.~\ref{fig:input states} to generate the states $\ket{\text{W}}$, $\ket{\text{G$^\prime$}}$, and $\ket{\text{GHZ$^\prime$}}$ upon post-selecting for three-fold coincidences at the three output ports. 
In order to verify the output states, we perform quantum state tomography and reconstruct the density matrices of the generated three-photon states (see Fig.~\ref{fig:Tomography}).

For the state $\ket{\text{W}}$, we obtain a fidelity of ${F_{\textrm{W}}=~0.873\pm0.011}$ and a purity of ${P_{\textrm{W}}=0.787\pm0.018}$.
The fidelity can be used as an entanglement witness to verify genuine tripartite entanglement~\cite{bourennane2004}. The witness is satisfied by a fidelity greater than 2/3, which the state $\ket{\text{W}}$ generated in our experiment fulfils.

For the state $\ket{\text{G$^\prime$}}$, we obtain a fidelity of ${F_{\textrm{G}^\prime}=0.834\pm0.011}$, and a purity of ${P_{\textrm{G}^\prime}=0.755\pm0.016}$. 
To verify the genuine tripartite entanglement of the state $\ket{\text{G$^\prime$}}$ we use an entanglement witness based on the overlap with the state $\ket{\text{GHZ$^\prime$}}$~\cite{bourennane2004}. An overlap greater than 0.5 indicates genuine tripartite entanglement for any state~\cite{bourennane2004}. The overlap measured in our experiment is $\bra{\text{GHZ}^\prime}\rho_{\text{G}^\prime}\ket{\text{GHZ}^\prime}=0.572\pm0.009$, confirming genuine tripartite entanglement~\cite{bourennane2004}.

For the state $\ket{\text{GHZ$^\prime$}}$, we obtain a fidelity of the generated state of ${F_{\textrm{GHZ}^\prime}=0.788\pm0.016}$, with a purity of ${P_{\textrm{GHZ}^\prime}=0.673\pm0.023}$. 
This fidelity surpasses the threshold of $0.5$, confirming the generation of genuine tripartite entanglement~\cite{bourennane2004}.
It furthermore surpasses the threshold of $0.75$, which is the maximal overlap of a general GHZ-type state with any W state. This proves that the generated state $\ket{\text{GHZ$^\prime$}}$ cannot belong to the class of W states, but belongs to the class of GHZ states~\cite{bourennane2004}. The errors in the fidelity values have been estimated using a Monte Carlo simulation.

Experimental imperfections lead to impurity and undesired imaginary terms in the density matrices, resulting in reduction in fidelity.
The main sources of error are a slight spectral distinguishability of the photons -- as verified from HOM experiments -- and higher-order photon emissions that, in the presence of losses, contaminate the density matrix components. We estimate the contribution of ratio of the six-photon emission to the four-photon emission to be on the order of $0.2\%$. Additional sources of error are polarisation rotations picked up during the transmission of the photons through the fibres that are not perfectly compensated for.
An average extinction ratio of 335:1 and 352:1 was obtained in the H/V basis and +/- basis, respectively. 

\section{Conclusion and Outlook}
In this work, we have demonstrated a scheme to generate photonic genuine tripartite entangled states belonging to different entanglement classes, i.e., the class of W states and the class of GHZ states, using the same experimental setup. The fidelities of the generated states confirm genuine tripartite entanglement in each case. Our approach is versatile, in the sense that it allows switching between the different states simply by applying local unitary operations at the inputs of a tritter. The entangled states are generated at $1550~\textrm{nm}$, which makes the implemented scheme suitable for networked quantum communication. 

Our scheme can be applied to any quantum protocols or applications that require a central server to provide entangled states. This server can then easily switch between different types of entangled states, possibly also randomly, providing additional security. The generated states form the basis for quantum protocols such as quantum secret sharing~\cite{hillery1999}, quantum conference key agreement~\cite{Grasselli2019}, and verifiable quantum random number generation~\cite{jacak2020}. Furthermore, the presented scheme can be used to herald different entangled states in matter systems mediated by photon interference, opening up new avenues for entangling matter nodes in quantum networks and quantum computing~\cite{maunz2007,van2022,raussendorf2001, raussendorf2003,gimeno2015}. 

Finally, it has been shown that the presented scheme can be extended to $N$-dimensions by using a $N\times N$ multiport beam splitter
and $N$ photons, and allows for the generation of arbitrary $N$-photon W and GHZ states~\cite{Lim2005,Bhatti2022}.

\begin{acknowledgments}
We acknowledge support from the Carl Zeiss Foundation, the Centre for Integrated Quantum Science and Technology (IQ$^\text{ST}$), the German Research Foundation (DFG), the Federal Ministry of Education and Research (BMBF, projects SiSiQ and PhotonQ), the Ministry of Economic Affairs, Labour and Tourism Baden-Württemberg (QORA), and the Federal Ministry for Economic Affairs and Energy (BMWi, project PlanQK).
\end{acknowledgments}

\end{document}